# Insertion of a magnesium(II)-octacarboranyl(hexylsulfanyl) porphyrazine into liposomes: a physico-chemical study

Anna Salvati,[1] Sandra Ristori,[1] Daniela Pietrangeli,[2] Julian Oberdisse,[3] Luca Calamai,[4] Giacomo Martini,[1] Giampaolo Ricciardi[2]*

[1]*University of Firenze, Department of Chemistry, Via della Lastruccia 3, 50019 Sesto Fiorentino (FI), Italy*

[2]*University of Basilicata, Department of Chemistry, Via N. Sauro 85, 85100 Potenza, Italy*

[3]*Laboratoire des Colloïdes, Verres et Nanomatériaux (LCVN), Université Montpellier II & CNRS, 34095 Montpellier, Cedex 05, France, and Laboratoire Léon Brillouin, CEA Saclay, 91191 Gif sur Yvette, France*

[4]*Department of Soil Science and Plant Nutrition, University of Florence, 50144 Florence, Italy*

*: Corresponding author: Prof. Giampaolo Ricciardi

Corresponding author:

Prof. Giampaolo Ricciardi, Dipartimento Di Chimica

Via N. Sauro, 85 , 85100 Potenza (Italy)

Tel: +39-0971-202142 Fax: +39-0971-202223

e-mail: rg010sci@unibas.it

**Abstract**

The synthesis, characterization and liposome insertion of a novel magnesium(II) carboranyl-porphyrazine, i.e. [2,3,7,8,12,13,17,18-octakis-(1,2-dicarba-*closo*-dodecaboranyl)-hexylthio-5,10,15,20-porphyrazine]magnesium(II) complex, MgHECSPz, is described. MgHECSPz was designed to improve the potentiality in multiple approach anticancer therapy. Liposomal formulations with different surface charge were prepared as delivering agents. The obtained loaded vectors were characterized by DLS, SAXS, SANS and ζ potential measurements in order to define the overall properties and structural details of loaded liposomes.

# 1. Introduction

A new approach for drug design must be based on the idea to combine multiple functionalities in one molecule or in its complex with the proper delivery vector. This way, it is possible to access different therapies with a single drug administration. Inside this scenario, Boron Neutron Capture Therapy (BNCT) [1-3] a two-step cancer treatment initially designed for radio-resistant highly invasive tumors, and for those not removable by surgery is nowadays acquiring a renovated potential. In particular, the BNCT strategy can be applied to pathologies such as gliomas in the brain and superficial cancers, coupled to Photodynamic and Photothermal Therapies (PDT and PTT, respectively) [4,5].

In order to be effective for BNCT, delivery of a large number of boron atoms is necessary, i.e. ~$10^9$ nuclei of $^{10}B$ per tumor cell [6]. The new generation of products investigated in the research on BNCT includes a large variety of carborane derivatives [7], where the icosahedral carborane cage has been linked to nucleic acids [8,9], sugars [10-12], porphyrins and related macrocycles [13-15].

In this paper, a novel magnesium(II) polycarboranylporphyrazine, i.e. [2,3,7,8,12,13,17,18-octakis-(1,2-dicarba-*closo*-dodecaboranyl)-hexylthio-5,10,15,20-porphyrazine]magnesium(II) (henceforth called MgHECSPz) (Figure 1) was synthesized and thoroughly characterized by conventional methods. The fact that carboranylporphyrazines may conjugate a high degree of peripheral carboranyl substitution with the peculiar electronic properties of the porphyrazinethiolate core motivated our choice. Indeed, these small-ring tetrapyrroles, depending on the nature of the peripheral substituents and the coordinated metal, show intense to very intense absorbance in the "therapeutic window", near-IR luminescence, or rapid radiationless decay of the primarily excited $S_1(Q)$ state [14,16].

Liposomes came into focus as valuable drug carriers and delivery systems because of their versatility and compatibility with respect to physiological environments [17]. In particular, for the specific application of porphyrins and related compounds [18], liposomes turned out to be capable of decreasing the tendency toward aggregation that strongly affects these photosensitizers and reduces their activity [19-21].

In this work liposome formulations with different surface charge were chosen for insertion of MgHECSPz in the lipid bilayer. The same long and flexible unsaturated dioleoyl chains were selected and used in all formulations in order to accommodate the bulky MgHECSPz molecule (~40 Å in its extended form). Plain and loaded vectors were characterized in term of structure and size by small angle neutron scattering (SANS), zeta potential ($\zeta$), and dynamic light scattering (DLS). Details at the bilayer level were obtained by small angle X-ray scattering (SAXS).

## 2. Materials and methods

*2.1. Materials*

All chemicals and solvents (Aldrich Chemicals Ltd.) used in the syntheses were of reagent grade and used as supplied. *o*-carborane was purchased from Ryskor Science, Inc. (USA). Solvents used in physical measurements were spectroscopic or HPLC grade. Diethyl ether and tetrahydrofuran (THF) were freshly distilled from sodium benzophenone ketyl under nitrogen. 1,2-dioleoyl-3-trimethylammonium-propane (chloride salt) (DOTAP, purity >99%), 1,2-dioleoyl-*sn*-glycero-3-phosphate (monosodium salt) (DOPA, purity >99%), and 1,2-dioleoyl-*sn*-glycero-3-phosphoethanolamine (DOPE, purity >99%), were purchased from Avanti Polar Lipids, Inc., Alabaster, AL, and used without further purification. 1,2-dioleoyl-*sn*-glycero-3-phosphocholine (DOPC, purity >99%) was purchased from Northern Lipids, Inc., Vancouver, CA.

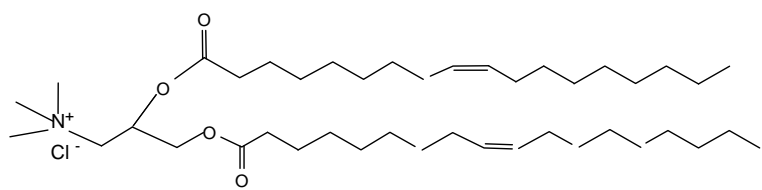

**DOTAP**

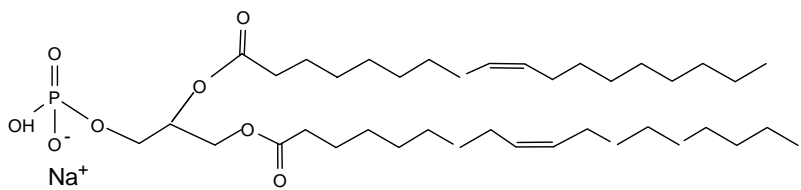

**DOPA**

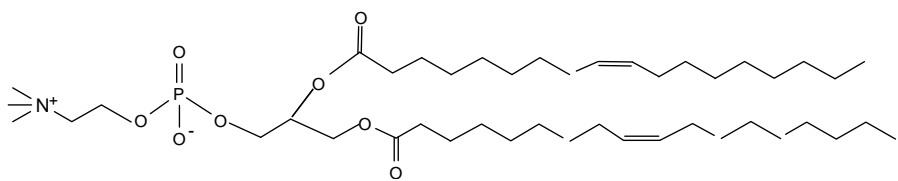

**DOPE**

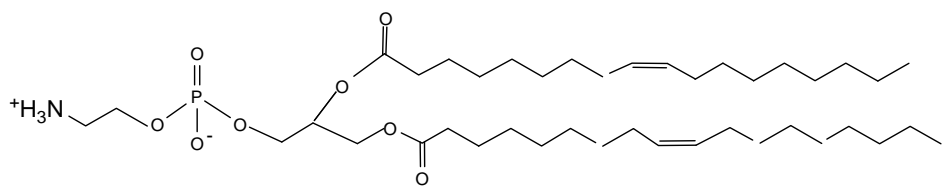

**DOPC**

*2.2. Liposome preparation*

Liposomes were prepared from a 1:1 molar ratio mixture of two di-oleoyl lipids, with different polar heads. One component was the helper lipid DOPE in all formulations. It has been shown that DOPE gives higher transfection efficiency, since it is able to form columnar inverted hexagonal, $H_{II}C$, structures, having higher fusogenic properties with respect to the lamellar structures ($L_\alpha$), usually formed by lipids with phosphocholine head groups (PC) [22-24]. The second lipid in the 1:1 mixture was varied among the zwitterionic DOPC, the cationic DOTAP, and the anionic di-oleoyl-glycero-phosphate DOPA in the form of monosodium salt. For each formulation, three different samples were prepared: 1) pure liposome; 2) liposome containing $2.1 \times 10^{-4}$ mol/L starting porphyrazine (MgHECSPz: lipid = 1 : 200); 3) liposome containing $4.2 \times 10^{-4}$ mol/L starting porphyrazine (MgHECSPz : lipid = 1 : 100). The total lipid content was $4.2 \times 10^{-2}$ mol/L in each sample.

Lipid dry powders were dissolved in chloroform and mixed in 1:1 ratio to obtain different liposomes formulations. The solvent was then evaporated to dryness under vacuum overnight. The resulting film was swollen at room temperature with MilliQ grade water. SANS samples were prepared in 99.9 % $D_2O$ (SIGMA). Upon vortexing, multilamellar vesicles were obtained and were submitted to eight cycles of freeze and thaw. Final downsizing and conversion to unilamellar vesicles was obtained by extrusion through 100 nm polycarbonate membranes with the LiposoFast apparatus, Avestin, Ottawa, CA. *3.2.* For the systems studied in this work, the employment of extrusion membranes with pore diameter smaller than 100 nm frequently led to the rupture of the membrane itself, with consequent lack of sample reproducibility. This was attributed to the impossibility of increasing the bilayer curvature above a certain extent in the experimental conditions (i.e. type of guest molecule, concentrations etc.) used.

*2.3. Methods*

$^1$H and $^{13}$C NMR data were recorded at 499.58 and 125.62 MHz, respectively, on an INOVA Varian 500 MHz spectrometer. $^{11}$B NMR data were recorded at 128.32 MHz on a Varian VNMRS-400 MHz spectrometer. Chemical shifts (δ) are expressed in parts per million (ppm). Coupling constants (*J*) are in Hz. The $^1$H and $^{13}$C NMR chemical shifts are relative to tetramethylsilane (TMS); peaks of the residual protons from the solvents were used as internal standards for $^1$H (δ 7.20 chloroform) and *all-d* solvent peaks for $^{13}$C (δ 77.0 chloroform); the $^{11}$B NMR chemical shifts are relative to external $BF_3 \cdot OEt_2$. All measurements were carried out at 298 K.

Infrared (IR) spectra were measured with a FT/IR-460-Plus JASCO spectrometer. Electronic absorption spectra were recorded using a UV-VIS-NIR Varian Cary 05E double-beam spectrophotometer. GC-MS spectra were measured with a Hewlett-Packard 6890 instrument.

Zeta potential measurements were performed with a Coulter DELSA 440 SX (Coulter Corporation, Miami, FL, USA). ζ was automatically calculated from the electrophoretic mobility by means of the Helmholtz-Smoluchowski relation. All samples were diluted 60 times to meet the instrumental sensitivity requirements. At least two different runs were carried out for each sample and all adducts showed a unimodal distribution of the ζ values.

Size measurements were performed by DLS with a Coulter Sub-Micron Particle Analyzer, Model N4SD, equipped with a 4 mW helium-neon laser (632.8 nm) and 90° detector. Mean diameter and polydispersity index, $(\mu/\gamma^2)^2$, of plain and drug-loaded liposomes were thus obtained.

Small Angle X-ray Scattering (SAXS) spectra were recorded on a high sensitivity pinhole camera with home built collimating optics and a rotating anode (Cu K$\alpha$) as the radiation source [25]. Samples were put into flat cells having 1 mm thickness and Nalophan$^{TM}$ windows. The accumulation time was 30 min. Details on the experimental setup and curve fitting are reported elsewhere [11].

Small Angle Neutron Scattering (SANS) experiments were performed on the research nuclear reactor ORPHEE, using the PAXY instrument of the Laboratoire Léon Brillouin (Saclay, France). Three different configurations were used ($\lambda$ = 12 Å, $D$ = 5 m; $\lambda$ = 4 Å, D = 5 m; $\lambda$ = 4 Å, D = 1m, where $\lambda$ is the neutron wavelength and $D$ the sample-detector distance). The overall $q$-range obtained was 0.0047–0.55 Å$^{-1}$. SANS diagrams were reduced to the absolute scale and the intensity at large angles (incoherent background) was subtracted before data analysis.

## 3. Results and discussion

### 3.1. Synthesis of MgHECSPz

MgHECSPz was synthesized with a five-steps procedure as shown in the following scheme:

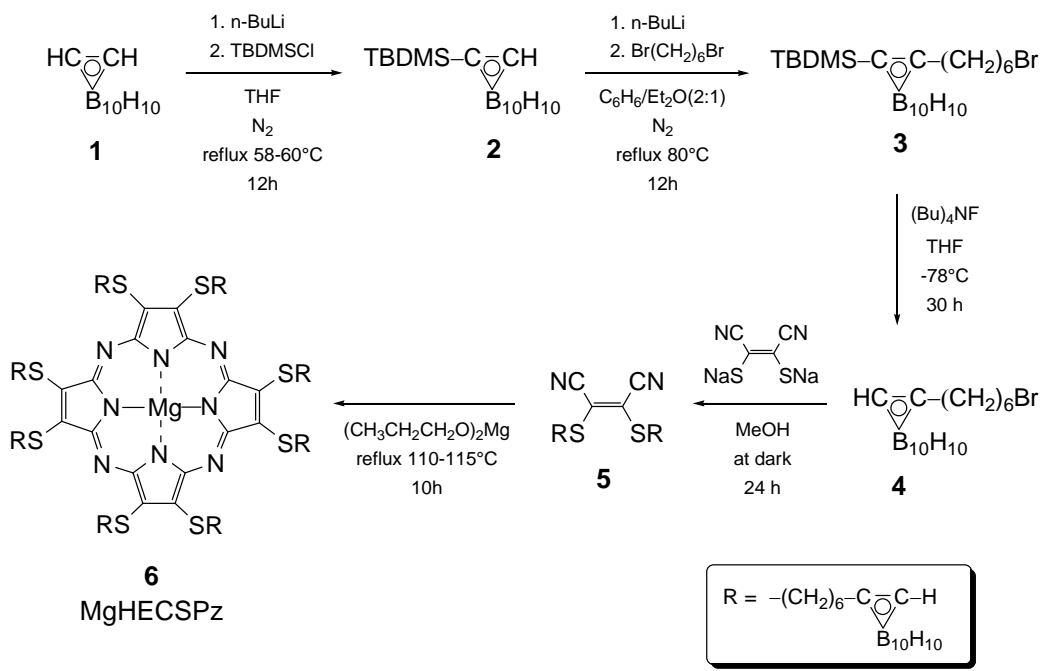

This procedure presented some modifications with respect to literature [14, 26,27]. More details on the synthetic procedure, the analytical characterization of the intermediates and on the final product will be published elsewhere. With these first three steps, the intermediates silyl-*o*-carborane (**2**), 6-(silyl-*o*-carboranyl)-1-bromohexane (**3**), and 6-(*o*-carboranyl)-1-bromohexane (**4**) were obtained in very good yields from 1,2-dicarba-*closo*-dodecaborane (**1**). Characterization data (mass spectra, $^1$H-, $^{13}$C, $^{11}$B-NMR, and IR spectra) of these compounds well agreed with similar compounds described in the literature [28,29]. The MgHECSPz precursor, the *cis*-2,3-bis[6-(1,2-*closo*-dodecarboranyl)hexylthio] maleonitrile, **5**, was obtained in ~20% yield by reaction of **4** with dimercaptomaleonitrile disodium salt at room temperature. Characterization data for **5**, including elemental analyses, were fully consistent with the expected structure confirming, in particular, that the *o*-carborane cages retain the *closo* structure in the compound. Cyclotetramerization of **5** on Mg$^{II}$(*n*-OPr)$_2$ (step 5) via a nucleophilic attack of the Mg$^{II}$(*n*-OPr)$_2$ on the *o*-carborane moieties proved to be the critical synthetic step. In this step degradation of the peripheral *o*-carboranes with formation of anionic (*nido*) species might occur [30]. However, we found that the impact of this process can be significantly attenuated by using mild reaction conditions (T < 115 °C) and a relatively short reaction time (~10h). Such procedure, though implying incomplete conversion of **5** (~ 30% of unreacted **5** can be retrieved), had the merit to reduce the formation of carborane-based byproducts, which could hardly be separated from the desired porphyrazine by chromatography.

Pure MgHECSPz was found to be insoluble in water either as monomer or in the form of aggregates.

The integrity of the eight *o*-carborane cages in the complex was confirmed by elemental analyses, and by $^1$H-, $^{13}$C-, $^{11}$B-NMR, ESI MS, and IR spectra. In particular, the $^{13}$C NMR spectrum showed two resonances at $\delta$ = 63.6 ppm and $\delta$ = 40.2 ppm which are the signature of the C-H and C-R (R = hexyl, in the actual case) carbons, respectively, belonging to a neutral *closo*-carborane cage [26,27]. In turn, the IR spectrum showed the symmetric and intense peak at 2588 cm$^{-1}$, attributable to the B-H stretching vibration of the *closo o*-carborane cage [27,29].

MgHECSPz exhibited the typical UV-Vis spectrum of alkylthioporphyrazines [16] in DMF, with relatively intense Q and B bands peaking at 671 nm and 384 nm, respectively, and a broad band due to $n_{sulfur} \to \pi^*$ transitions in between (502 nm). The optical spectrum of MgHECSPz solubilized in weakly coordinating solvents, such as chloroform (Figure 2), showed a clearly split Q-band and a blue shifted B-band. This suggested that the complex had a marked tendency to give dimers or even larger aggregates in hydrophobic environments, most likely by setting up effective Mg$^{...}$S intermolecular interactions.

Aggregation could be promptly disrupted, however, upon addition of a minimal amount (<1% v/v) of a coordinating solvent, such as methanol or pyridine (Figure 2). Quite interestingly, the "monomeric" form of the complex could also be restored upon addition of an equimolar amount of phospholipid. This behavior, which was in line with the well documented tendency of $Mg^{II}$ porphyrazines to axially coordinate even relatively weak σ-donor ligands, such as water [29], suggested that the complex, when incorporated in the liposomal matrix, set up interactions with the polar heads of the lipids, particularly with the phosphate groups. This finding was also in agreement with the behavior of similar non metallated tetra-aza macrocycles [14].

*3.2. Loding of MgHECSPz into liposomes*

MgHECSPz was added in the desired concentration from a $2 \times 10^{-3}$ M stock solution in $CHCl_3$ to the mixed lipid solution in the same solvent. Porphyarzine containing liposomes then were prepared following the same procedure described in section 2.2 for plain liposomes.

The rate of MgHECSPz loading was determined as described earlier [14]. It was found to be in the range 6-12% of added MgHECSPz, independently of lipid formulation and starting porphyrazine content.

*3.3. Characterization of loaded liposomes*

*A) Zeta potential analysis.* Figure 3 shows the Zeta Potential values of the different liposome systems as a function of nominal porphyrazine concentration. Pure DOTAP/DOPE liposomes had a Zeta Potential value of +45(±5) mV, as expected for the presence of the cationic lipid DOTAP. On the other hand, the Zeta Potential was

–65(±10) mV for pure DOPA/DOPE, because of the negatively charged DOPA. For DOPC/DOPE liposomes the measured ζ value was –5(±1) mV. This meant that the zwitterionic liposomes used in this study behaved as slightly negatively charged objects in water solution and that, at molecular level, the phosphate groups were more exposed than the choline moieties on the outer surface of liposomes.

In MgHECSPz-DOTAP/DOPE systems the Zeta Potential decreased with increasing porphyrazine concentration, while it increased in DOPA/DOPE and DOPC/DOPE series. This confirmed the insertion of the porphyrazine in all liposome formulations. At the same time either the decrease or the increase of ζ suggested that MgHECSPz (not charged) was able to dilute the liposome surface charge upon insertion in the bilayer region. In particular, for the DOPC/DOPE series, Zeta Potential inverted its sign by increasing the porphyrazine content from the slightly negative value of pure liposomes toward positive values. For this formulation, the results were more variable with respect to the other type of liposomes, and sometimes negative values were obtained for both plain and porphyrazine-loaded liposomes.

Table 1 shows the diameter and the polydispersity index of the same systems reported in Figure 3, as measured by dynamic light scattering.

Pure liposome diameters were slightly larger than 100 nm, as expected from the preparation procedure. In the DOTAP/DOPE and DOPA/DOPE series, liposome size and polydispersity slightly increased with increasing porphyrazine concentration. For DOPC/DOPE liposomes the insertion of MgHECSPz induced a stronger effect, if compared with the other formulations. This difference was in line with the trends described for Zeta Potential values. The peculiarity of size growth observed for DOPC/DOPE formulation was related to the properties of the phosphocholine head group, which presents a stronger tendency to stack into multilamellar structures with respect to other types of polar head. The insertion of the porphyrazine ring enhanced this tendency, as confirmed for similar molecules in the same lipid formulation [14].

*B) SAXS analysis.* Figure 4a shows the evolution of SAXS profiles for plain and MgHECSPz-loaded DOPA/DOPE liposomes as a function of the porphyrazine starting concentration. To fit these curves, the form factor of a bilayer was used:

$$I(q) = \frac{4\pi}{q^4} \Sigma \, [(\rho_{c,e} - \rho_{h,e})\sin(q\frac{t_c}{2}) + (\rho_{h,e} - \rho_{s,e})\sin(q\frac{2t_h + t_c}{2})]^2 \qquad (1)$$

where $\rho_c$, $\rho_h$, $t_c$ and $t_h$ are the electronic density and the thickness of the hydrophobic and hydrophilic layer, respectively; $\rho_s$ is the electronic density of the solvent and $\Sigma$ the interface extension per unit volume [31]. A Gaussian distribution of the bilayer hydrophobic thickness (with half-width at half maximum $\delta t_c$) allowed to take into account a certain degree of polydispersity. The corresponding fittings are reported in Figure 4b-d and the best-fit parameters are listed in Table 2. A trend could be evidenced and it was consistent with an increase in bilayer thickness and polydispersity at increasing porphyrazine content. In particular, a sort of bilayer swelling upon drug addition occurred, probably due to the bulky structure of the porphyrazine used in this study. The above results were also in agreement with those obtained by Zeta Potential and DLS measurements, which gave an analogous trend for the overall properties (i.e. surface charge and mean size) of plain- and porphyrazine-loaded liposomes.

Contrarily to what happened for negative liposomes, in the cationic DOTAP/DOPE series the bilayer did not appear to be influenced by porphyrazine at the lipid/MgHECSPz loading ratios investigated in this works (data not shown). In fact, the differences between the SAXS diagrams of plain- and MgHECSPz-loaded liposomes were superimposed within the limits of experimental error. A unique fitting curve reproduced the general trend. The corresponding best-fit parameters are reported in Table 2.

The SAXS profiles of the zwitterionic liposomes, with and without added MgHECSPz are reported in Figure 5. These systems showed a peculiar behaviour with respect to either positively or negatively charged starting liposomes, because the insertion of the porphyrazine molecules induced partial formation of multilamellar vesicles.

The presence of these latter structures was clearly established by the appearance of Bragg peaks, superimposed to the diffuse scattering curve of monolamellar liposomes. In particular, the first two reflections of the (l00) lamellar stacking were detected, from which the interlamellar distance $d = 2\pi/q_0$ could be calculated ($q_0$ is the position of the first order Bragg peak). The formation of pauci-lamellar structures in extruded liposomes has been reported in the literature. For instance, Kucerka et al. [32] have studied and modeled the extent of this phenomenon as a function of vescicle size and bilayer net charge by doping DOPC liposomes with the negative lipid DOPS.

The value obtained for interlamellar distance in the present systems (61.5 ± 0.5 Å) was in agreement with the literature [33]. The occurrence of partial stacking in DOPC/DOPE liposomes was also in agreement with the high polydispersity values found for the vesicle overall size by DLS measurement, as reported in Table 1. This effect was ascribed to the well known tendency of porphyrin-like macrocycles to stack on top of each other, especially when in the presence of water or another polar solvent. In these conditions porphyrins and related macrocycles tend to form dimers or larger columnar aggregates, as widely reported in the literature [20,21] and as discussed in section 3.1 of this paper.

Pure DOPC/DOPE liposomes were monolamellar and their SAXS profile was fitted with the usual procedure followed in this work. The corresponding best-fit parameters are listed in Table 2. More detailed information on the fine structure of systems where monolamellar and pauci-lamellar structures coexist in solution could be obtained by extensive SAXS data treatment and curve fitting [32,34]. In particular, Nallet et al. [34] have obtained the bilayer elastic constants, as well as an estimation of the number of lamellae, in binary lyotropic smectics, which show diffuse scattering underlying Bragg peaks. These authors have built up a model that takes into account both the membrane geometry and the thermodynamic origin of the layer displacement fluctuations.

.

*C) SANS analysis.* The experimental SANS patterns of plain and MgHECSPz-loaded liposomes were fitted by using the complete form factor of monolamellar vesicles $P_{ves}$, with a homogeneous bilayer contrast:

$$P_{ves} = \left[ \frac{4\pi}{3} (R+t)^3 f(q(R+t)) - \frac{4\pi}{3} R^3 f(qR) \right]^2 \qquad (2a)$$

with *f(q(R+t))* and *f(qR)* in the form:

$$f(x) = \frac{sinx - xcosx}{x^3} \tag{2b}$$

where $R$ is the inner radius of the vesicle and $t = 2t_h + t_c$ the total bilayer thickness. Note that, unlike SAXS, SANS probes mainly the hydrophobic core region, whereas the hydrated head groups contribute much less to the scattering. The scattered intensity $I(q)$ then reads:

$$I(q) = \frac{n}{V} \Delta\rho^2 \, P_{ves} \tag{3}$$

where $\Delta\rho$ is the average neutron scattering length density difference between the vesicle components and the deuterated solvent (SANS contrast), $n/V$ is the number density of scattering objects, i.e. the number of vesicles per unit volume, which is directly related to the volume fraction for vesicles of a given size.

All fits were carried out by using the form factor in eq. (2), integrated numerically over a Gaussian distribution function in radius (with standard deviation $\sigma_R$), in thickness (with standard deviation $\sigma_t$), and convoluted with the resolution function of the instrument [35,36].

In the case of SANS, the theoretical shape of vesicle scattering curve (Eqs. 2-3) would be the following [37]: at small $q$ values, there should be no relevant contribution from the interaction among different vesicles, given the low concentrations used in this work, and the intensity should start from a value proportional to the total surfactant mass. As $q$ increases, the intensity decreases proportionally to $\exp(-q^2R_g^2/3)$, where $R_g$ is the radius of gyration of the vesicle, i.e. roughly the radius of these thin shell objects. At intermediate $q$, the intensity is proportional to $q^{-2}$, which is the characteristic power law for bilayer scattering. This was what we observe in Figure 6 and 7, where no Guinier plateau was found at low *q*, indicating that the vesicles were too large for our experimental *q*-range. The radius measured in DLS was thus taken as an input parameter for the fitting although, as explained below, this did not necessarily produced the best fit. At high *q*, a second power law was found, proportional to $q^{-4}$, which is the Porod scattering from the surface.

Figure 6 shows the SANS curve of DOTAP/DOPE liposomes and the corresponding best fit. As in the case of SAXS, no large differences could be detected between plain and porphyrazine-loaded DOTAP/DOPE liposomes. The fitting was carried out with the parameters listed in Table 3. In particular, the radius value was obtained from DLS measurements and the thickness value was 35 Å with a half-width at half maximum of $\sigma_t$ = 3 Å for the Gaussian distribution. This size and degree of polydispersity reproduced the features of experimental curves at large *q*-values, e.g. the 'kink' located at ≈ $2\pi/35$ Å, in a satisfying manner. Contrarily to the SAXS analysis, which concentrated on the bilayer profile, the lower-*q* data of the SANS investigation allowed to capture the $q^{-2}$ power law which is characteristic of large-scale bilayer scattering, as mentioned above.

Essentially the same analysis could be carried out for plain and MgHECSPz loaded DOPC/DOPE liposomes. The corresponding fitting values are also reported in table 3. It is to be noted that in the case of loaded DOPC/DOPE liposomes, SANS patterns did not show the presence of pauci-lamellar structures, which were evidenced by SAXS curves as shown in Figure 5. This suggested that the amount of this kind of structures was limited, and that only the SAXS sensitivity toward the bilayer properties could reveal their presence.

The SANS curves of negatively charged DOPA/DOPE liposomes showed a slightly different behavior (Figure 7), since at low-q, an evolution in the large-scale structure of vesicles was found. Although the fitting could be done with the same procedure used for cationic and zwitterionic liposomes, that is by taking the outer radius from light scattering measurement, the corresponding fitting gave poor reproduction of the experimental data in the low-q region. This pushed us to choose a different approach, that is let the vesicle size be varied instead of being constrained to the value obtained by DLS measurements .

For pure DOPA/DOPE, the SANS curve showed almost no structure at low angles. Fitting this curve with the radius taken from DLS (600 Å, $\sigma_R$ = 60 Å), allowed to obtain some low-q oscillations not visible in the SANS-data and it was necessary to increase the radius till about 1000 Å and introduce a high polydispersity in radius (200 Å), in order to obtain scattering curves compatible with the experimental data, as the one shown in Figure 7b. Note that we fixed the volume fraction to the value used for the SAXS fitting ($\phi$ = 2.5%, cf. Table 2). The prefactor of the intensity was then governed by the average contrast of the bilayer, cf. eq.(3), and in the fit of Figure 6b $\Delta\rho$ = 6.6 $10^{10}$ cm$^{-2}$ was used. This corresponded to a scattering length density of – 0.2 $10^{10}$ cm$^{-2}$ of the bilayer core, in good agreement with previous results [11]. As 2 $10^{-4}$ M of MgHECSPz was added, the intensity curve showed some very weak oscillations at low-q. Again, as we did not have access to sufficiently low-q data, in order to reach the Guinier regime, we first tried the radius obtained from DLS (650 Å, $\sigma_R$ = 60 Å), which also predicted weak oscillation, but considerably out of phase with respect to the observed one. Decreasing the inner radius to 535 Å with, as before, a rather large polydispersity in radius (120 Å), we could reproduce the scattered intensity rather well, apart from a slight overestimation of the intensity at low angles (Figure 7c). As in the case of the pure DOPA/DOPE sample, we kept the same volume fraction used for the SAXS treatment ($\phi$ = 2.7%, cf. Table 2), and adjusted the bilayer contrast for a good fit ($\Delta\rho$ = 6.4 $10^{10}$ cm$^{-2}$). This corresponded to a difference of less than five percent with respect to pure DOPA/DOPE, which is an adequate deviation, since the commonly accepted error is 5 to 10 % in determining the absolute units. We should also mention that the insertion of MgHECSPz corresponded to calculated undetectably small change in the scattering length density of the bilayer (less than 1%).

The DOPA/DOPE data set with starting 4x10$^{-4}$ M of MgHECSPz also showed an oscillation at low q, which was not reproduced by using the radius obtained from DLS (750 Å, $\sigma_R$ = 60 Å). On the contrary, a lower (inner) radius of 485 Å with a polydispersity in radius of $\sigma_R$ = 80 Å yielded the fitting shown in Figure 7d. Again the low-q intensity was slightly too high, but the position of the oscillations and the general shape of the curve could be reproduced in a satisfactory manner. As with the previous two data sets, the volume fraction was fixed ($\phi$ = 2.8%, cf. Table 2), and a bilayer contrast of $\Delta\rho$ = 6.4 $10^{10}$ cm$^{-2}$ was used for the fit plotted in Figure 6d.

As it can be seen from the values reported in Table 3, the SANS patterns did not change significantly at intermediate and high q within each series. This indicated that no relevant effect could be evidenced by SANS upon porphyrazine insertion in the liposome bilayers. The thickness of the bilayers was slightly smaller than the value obtained from SAXS measurements and this was ascribed to the hydration of the bilayer, which brings about differences in the contrast for the two small angle scattering techniques [38]. In summary, for the DOPA/DOPE series, vesicles seemed to shrink with porphyrazine-loading, and become more monodisperse. One may note that the scattered intensities systematically overestimate the intensity in this q-region. This could be the signature of some weak interaction between vesicles, i.e. the intensity may be suppressed by the inter-vesicle structure factor. If we estimate the order of magnitude, vesicles of radius 600 Å at 3% are typically at a center-to-center distance of 1750 Å. This was just outside our window of observation ($2\pi/1750$ Å = 3.6 $10^{-3}$ Å$^{-1}$), and it is possible that weak repulsion among vesicles induced some suppression in the low-q scattering.

## 4. Conclusions

In this paper, we described characterization of a magnesium(II) complex of a novel polycarboranylporphyrazine, MgHECSPz, and its insertion in different liposome formulations and the resulting adducts.

Insertion was reported by the change in several physico-chemical parameters of liposomes, e.g. their size and polydispersity, Zeta Potential and bilayer thickness. The overall effect observed upon loading was a small change in the liposome size and polydispersity, as well as a dilution of the charge. This latter factor agreed with the trend of the Zeta Potential values in the different systems. SAXS and SANS analysis showed that only the anionic DOPA/DOPE liposomes exhibited a definite trend of varying size and bilayer thickness with increasing MgHECSPz content.

The most relevant result of the present study was that in all cases the structure and bilayer properties of loaded liposomes were not drastically changed. This is a favorable starting point for possible applications of these systems in the biomedical field.

**Aknowledgements**

We thank dr. Olivier Spalla and dr. Olivier Taché, DSM, CEA Saclay, for help in SAXS measurements and analysis. The Laboratoire Leon Brillouin (LLB, Saclay) is acknowledged for beam time allocation in the framework of the NM13 European Program. The Italian Consorzio per le Superfici e per le Grandi Interfasi (CSGI) and the University of Florence are acknowledged for financial support.

**Table 1.** Diameter and polydispersity index of the investigated liposomes.

| Sample | Diameter (nm) | Polid. index |
|---|---|---|
| Pure DOTAP/DOPE | 112(±4) | 0.045 |
| DOTAP/DOPE + MgHECSPz = $2.1 \times 10^{-4}$ M | 120(±5) | 0.079 |
| DOTAP/DOPE + MgHECSPz = $4.2 \times 10^{-4}$ M | 126(±5) | 0.073 |
| Pure DOPC/DOPE | 120(±4) | 0.090 |
| DOPC/DOPE + MgHECSPz = $2.1 \times 10^{-4}$ M | 218(±8) | 0.43 |
| DOPC/DOPE + MgHECSPz = $4.2 \times 10^{-4}$ M | 190(±7) | 0.41 |
| Pure DOPA/DOPE | 117(±4) | 0.093 |
| DOPA/DOPE + MgHECSPz = $2.1 \times 10^{-4}$ M | 150(±5) | 0.20 |
| DOPA/DOPE + MgHECSPz = $4.2 \times 10^{-4}$ M | 150((±5) | 0.29 |

**Table 2.** Best fit parameters for plain- and porphyrazine-loaded liposomes.

| Sample | $\phi$ | $\rho_c$ (cm$^{-2}$) | $\rho_h$ (cm$^{-2}$) | $t_c$ (Å) | $\delta t_c$ (Å) | $r_h$ (Å) | $t_{bil}$ (Å) |
|---|---|---|---|---|---|---|---|
| Pure DOTAP/DOPE | 0.025 | 7.6×10$^{10}$ | 1.32×10$^{11}$ | 28.0 | 0.5 | 5.4 | 39.6 |
| Pure DOPC/DOPE | 0.025 | 7.6×10$^{10}$ | 1.26×10$^{11}$ | 27 | 1.0 | 7.5 | 42 |
| Pure DOPA/DOPE | 0.025 | 7.6×10$^{10}$ | 1.38×10$^{11}$ | 25.0 | 0.3 | 6.0 | 39.0 |
| DOPA/DOPE + MgHECSPz = 2.1×10$^{-4}$ M | 0.027 | 7.5×10$^{10}$ | 1.41×10$^{11}$ | 27.6 | 0.5 | 6.5 | 40.6 |
| DOPA/DOPE + MgHECSPz = 4.2×10$^{-4}$ M | 0.028 | 7.2×10$^{10}$ | 1.43×10$^{11}$ | 28.3 | 0.6 | 7.2 | 42.7 |

**Table 3.** Fit parameters for SANS curves.

| Sample | Inner radius (Å) | $\sigma_R$ (Å) | Bilayer thickness (Å) ($\sigma_t = 3$ Å) |
|---|---|---|---|
| Pure DOTAP/DOPE | 600 | 50 | 35 |
| DOTAP/DOPE + MgHECSPz 2.1×10$^{-4}$ M | 600 | 60 | 35 |
| DOTAP/DOPE + MgHECSPz 4.2×10$^{-4}$ M | 650 | 60 | 35 |
| Pure DOPC/DOPE | 600 | 50 | 34 |
| DOPC/DOPE + MgHECSPz 2.1×10$^{-4}$ M | 1000 | 60 | 34 |
| DOPC/DOPE + MgHECSPz 4.2×10$^{-4}$ M | 1000 | 80 | 35 |
| Pure DOPA/DOPE | 1000[a] | 200[a] | 35 |
| DOPA/DOPE + MgHECSPz 2.1×10$^{-4}$ M | 535[a] | 120[a] | 35 |
| DOPA/DOPE + MgHECSPz 4.2×10$^{-4}$ M | 485[a] | 80[a] | 35 |

[a] These data have been obtained with a slightly different simulation procedure, as detailed in the text.

**Captions to figures**

Figure 1. Molecular structure of [2,3,7,8,12,13,17,18-octakis-(1,2-dicarba-*closo*-dodecaboranyl)-hexylthio-5,10,15,20-porphyrazine]magnesium(II), MgHECSPz. Hydrogen atoms are omitted for the sake of clarity.

Figure 2. UV-Visible spectrum of MgHECSPz ($6.2 \cdot 10^{-6}$M) at room temperature in $CHCl_3$ (thin line) and $CHCl_3$/MeOH (99:1 v/v) (tick line).

Figure 3. Zeta Potential of DOTAP/DOPE (A), DOPA/DOPE (B) and DOPC/DOPE (C) liposomes as a function of the nominal content of MgHECSPz.

Figure 4. SAXS curves of DOPA/DOPE series. (a) the three experimental curves: black squares: pure liposomes; red squares: liposomes with $2.1 \times 10^{-4}$ M MgHECSPz; green squares: liposomes with $4.2 \times 10^{-4}$ M MgHECSPz.; (b-d) the experimental (square symbols as above) and the corresponding fitting (solid lines) curves for plain (b) and loaded (c, d) samples.

Figure 5. Experimental SAXS curves and fitting of DOPC/DOPE liposomes. Black squares: pure liposomes; red squares: liposomes with $2.1 \times 10^{-4}$ M MgHECSPz; green squares: liposomes with $4.2 \times 10^{-4}$ M MgHECSPz. Solid line: fitting curve.

Figure 6. Experimental SANS curve (empty squares) and fitting (filled circles) of pure DOTAP/DOPE liposomes.

Figure 7. SANS curves of DOPA/DOPE series. (a) the three experimental curves: black squares: pure liposomes; red squares: liposomes with $2 \times 10^{-4}$ M MgHECSPz; green squares: liposomes with $4 \times 10^{-4}$ M MgHECSPz.; (b-d) the experimental (square symbols, as above) and the corresponding fitting (filled circles) curves for plain (b) and loaded (c, d) vectors.

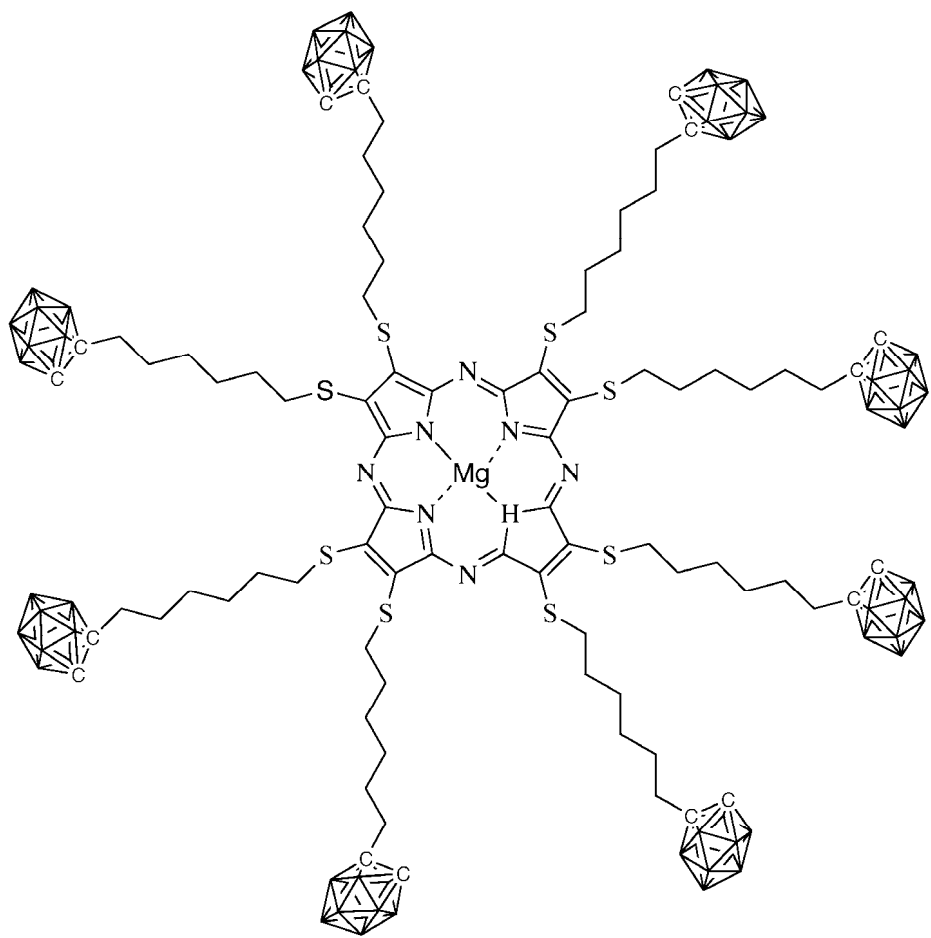

Figure 1. Molecular structure of [2,3,7,8,12,13,17,18-octakis-(1,2-dicarba-*closo*-dodecaboranyl)-hexylthio-5,10,15,20-porphyrazine]magnesium(II), MgHECSPz. Hydrogen atoms are omitted for the sake of clarity.

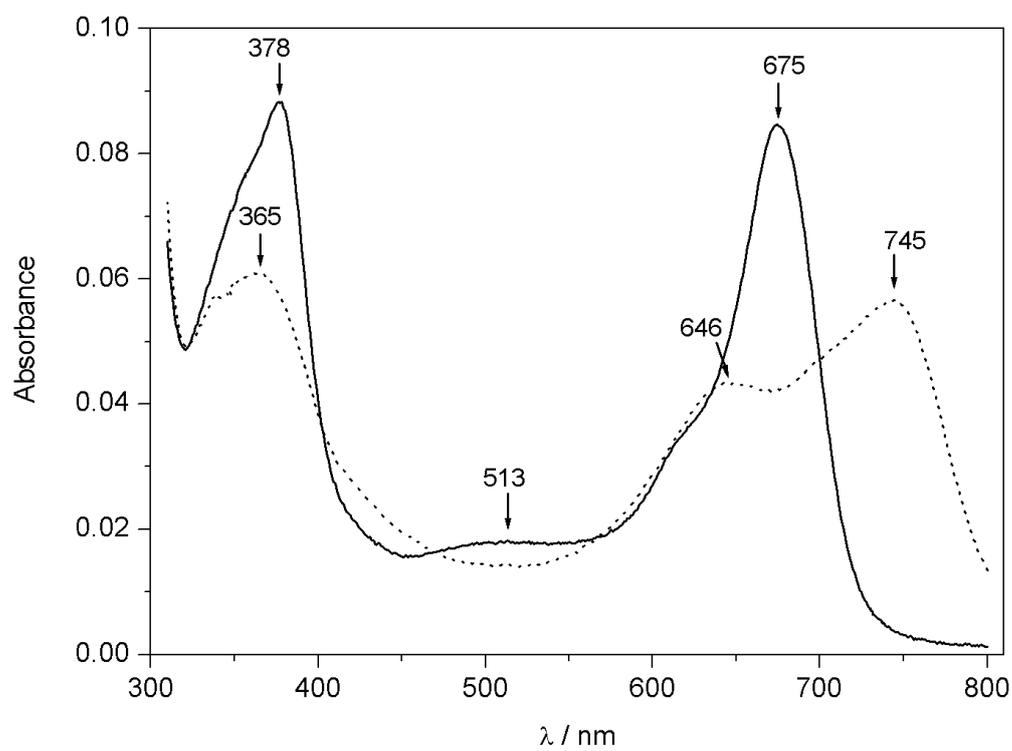

Figure 2. UV-Visible spectrum of MgHECSPz (6.2·10$^{-6}$M) at room temperature in CHCl$_3$ (dashed line) and CHCl$_3$/MeOH (99:1 v/v) (full line).

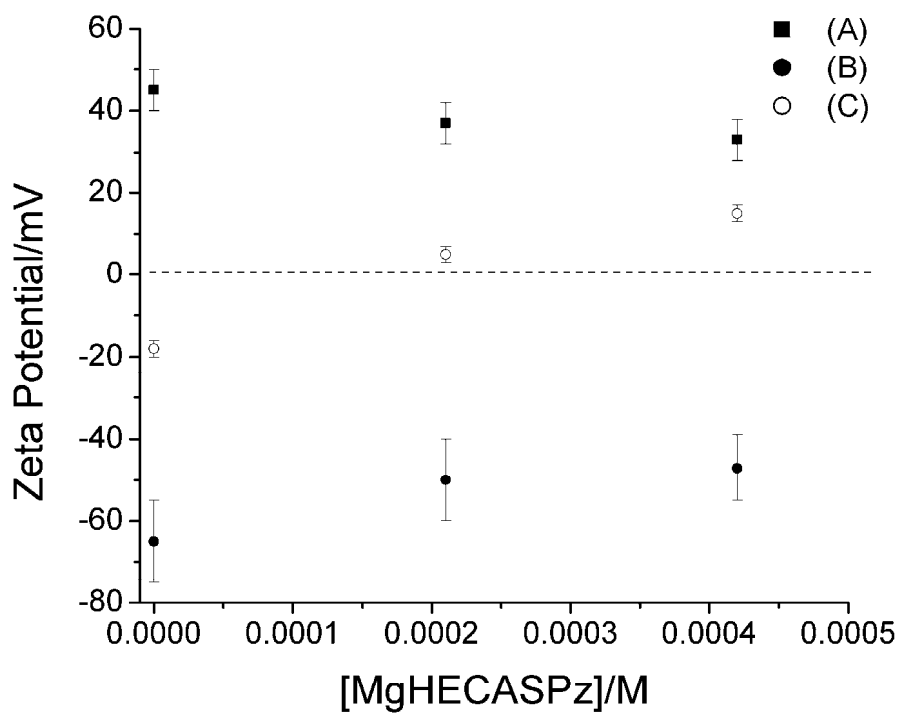

Figure 3. Zeta Potential of DOTAP/DOPE (A), DOPA/DOPE (B) and DOPC/DOPE (C) liposomes as a function of the nominal content of MgHECSPz.

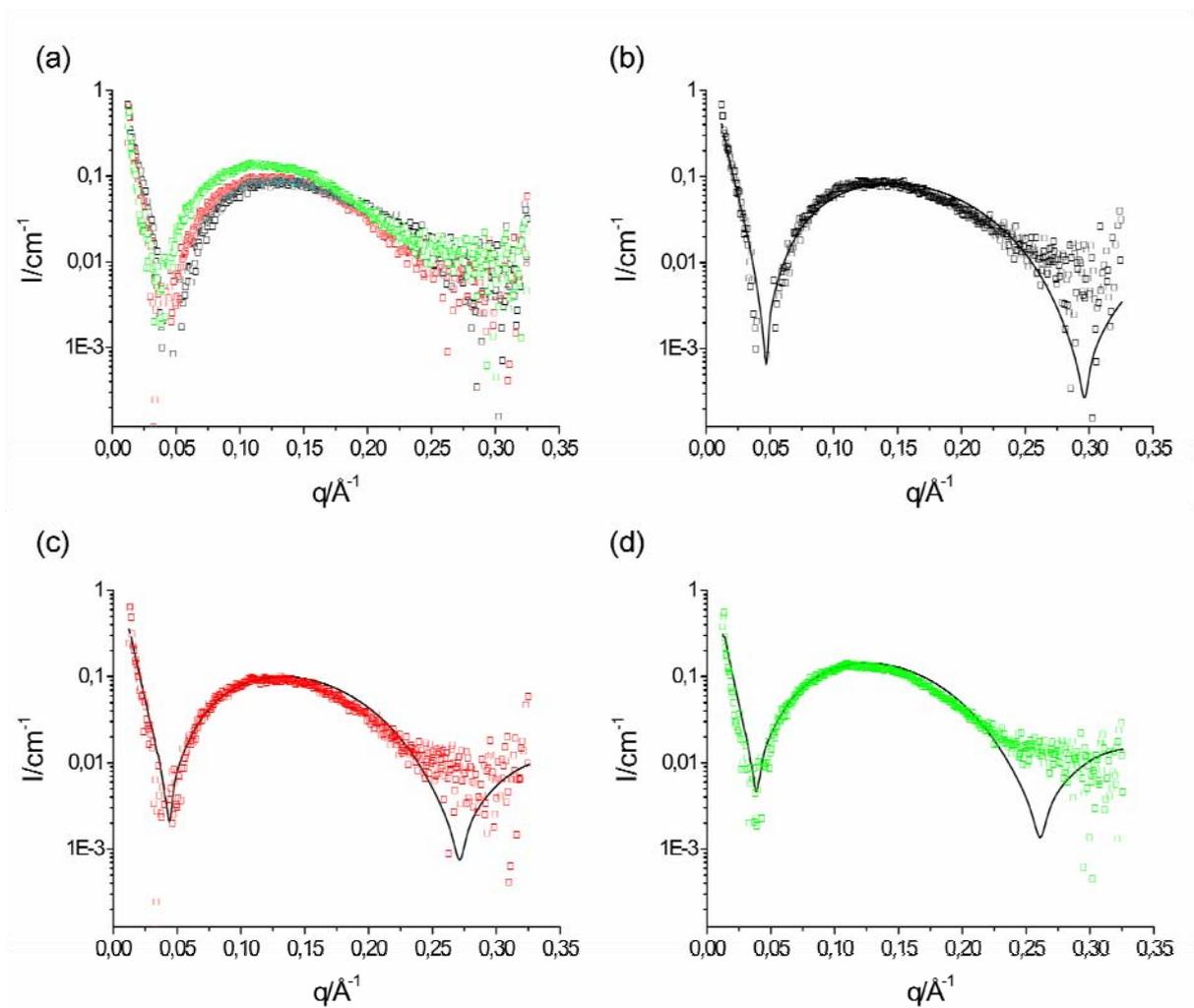

Figure 4. SAXS curves of DOPA/DOPE series. (a) the three experimental curves: black squares: pure liposomes; red squares: liposomes with $2.1\times10^{-4}$ M MgHECSPz; green squares: liposomes with $4.2\times10^{-4}$ M MgHECSPz.; (b-d) the experimental (square symbols as above) and the corresponding fitting (solid lines) curves for plain (b) and loaded (c, d) samples.

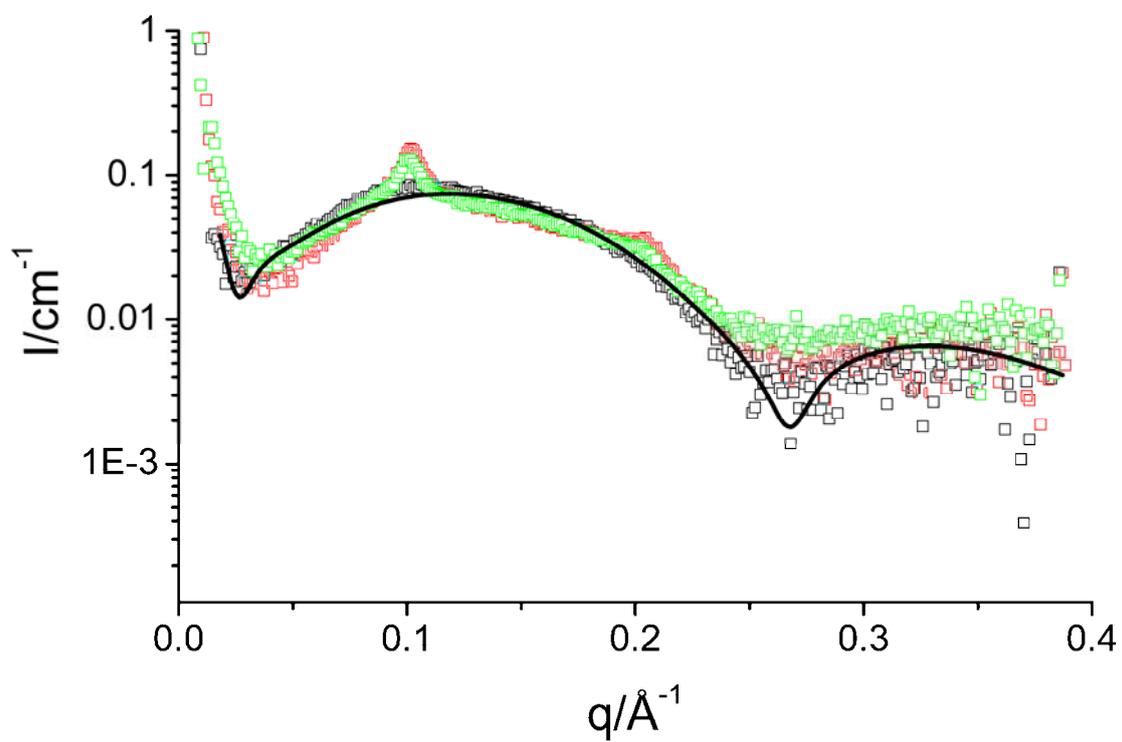

Figure 5. Experimental SAXS curves and fitting of DOPC/DOPE liposomes. Black squares: pure liposomes; red squares: liposomes with 2.1×10$^{-4}$ M MgHECSPz; green squares: liposomes with 4.2×10$^{-4}$ M MgHECSPz. Solid line: fitting curve.

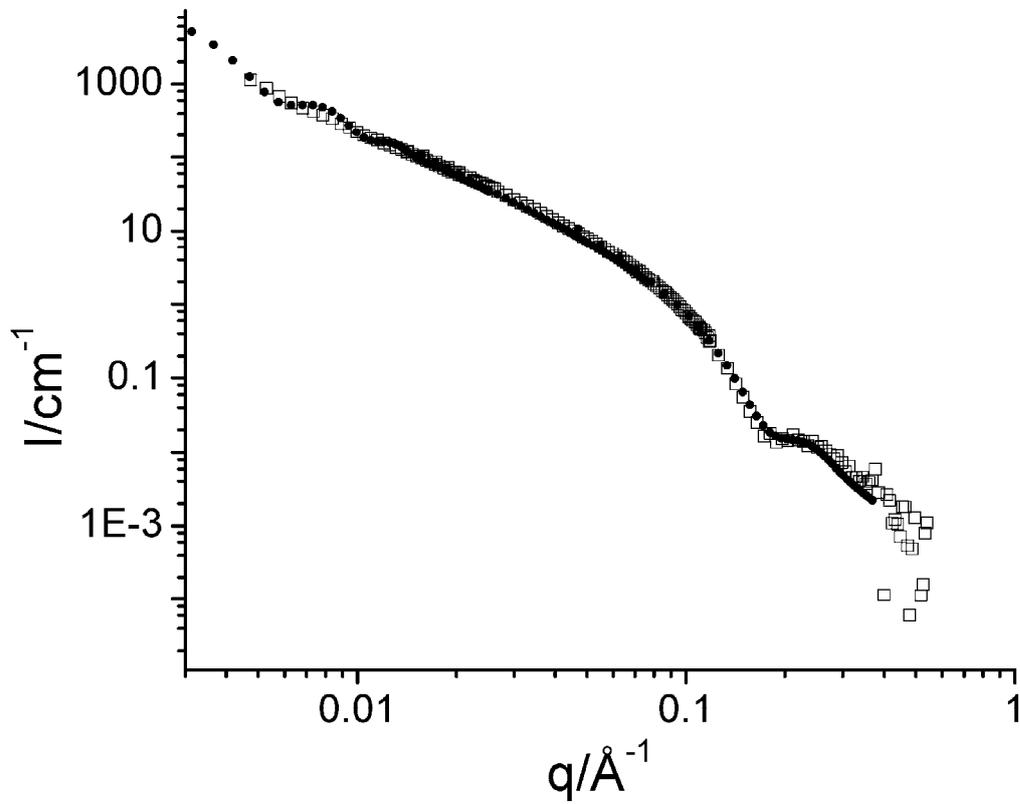

Figure 6. Experimental SANS curve (empty squares) and fitting (filled circles) of pure DOTAP/DOPE liposomes.

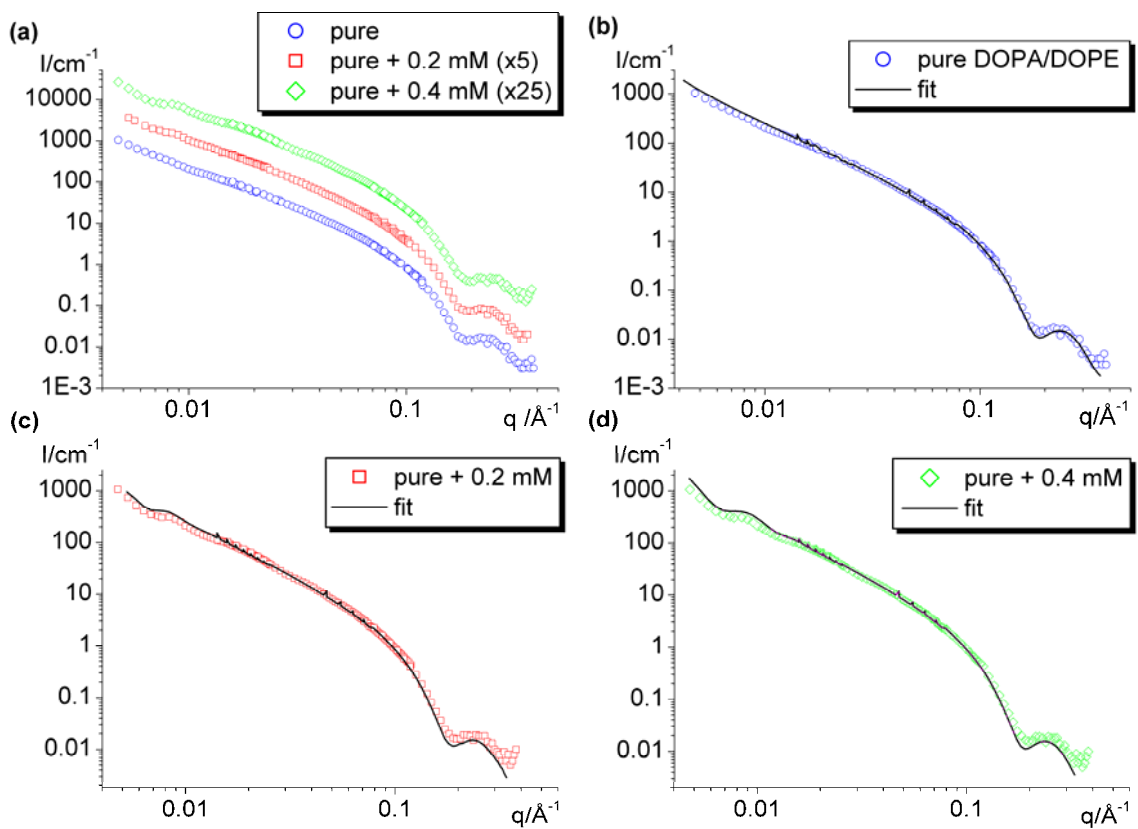

Figure 7. SANS curves of DOPA/DOPE series. (a) the three experimental curves: black squares: pure liposomes; red squares: liposomes with 2×10$^{-4}$ M MgHECSPz; green squares: liposomes with 4×10$^{-4}$ M MgHECSPz.; (b-d) the experimental (square symbols, as above) and the corresponding fitting (filled circles) curves for plain (b) and loaded (c, d) vectors.